\documentclass[aps,prd,twocolumn,showpacs]{revtex4-1}

\usepackage{graphicx}
\usepackage{dcolumn}
\usepackage{bm}
\usepackage{pdfpages}

\begin{document}

\title{Thermodynamic consistency, quark mass scaling,
       and properties of strange matter}
\author{C.~J.~Xia,$^1$ G.~X.~Peng,$^{1,2}$ S.~W.~Chen,$^{1,3}$
        Z.~Y. Lu,$^1$ and J.~F. Xu$^1$
        }
\affiliation{%
$^1$\mbox{School of Physics,
    University of Chinese Academy of Sciences, Beijing 100049, China}\\
$^2$\mbox{Theoretical Physics Center for Science Facilities,
    Institute of High Energy Physics,  Beijing 100049, China}\\
$^3$China Information Technology Security Evaluation Center,
   Shangdixilu 8, Beijing 100085, China
            }

\begin{abstract}
The previous thermodynamic treatment for models with density and/or
temperature dependent quark masses is shown to be inconsistent
with the requirement of fundamental thermodynamics.
We therefore study a fully self-consistent one
according to the fundamental differential equation of thermodynamics.
After obtaining a new quark mass scaling with the inclusion of
both confinement and leading-order perturbative interactions,
we investigate properties of strange quark matter in the fully consistent
thermodynamic treatment. It is found that the equation of state become stiffer,
and accordingly, the maximum mass of strange stars is
as large as about 2 times the solar mass, if strange quark matter
is absolutely or metastable.
\end{abstract}

\pacs{21.65.Qr, 05.70.Ce, 12.39.-x}

\maketitle

\section{Introduction}
\label{intro}

Since its possible absolute stability was conjectured nearly
thirty years ago \cite{Witten1984}, strange quark matter (SQM) has
been playing an important role in many interesting fields, for example,
the deconfinement phase transition \cite{Peng2008,Wen2013,Zheng2014,Burgio2002,Logoteta2013,Orsaria2014},
the hot and dense matter in heavy ion collisions \cite{Wiedemann2013},
the structure of compact stars \cite{Bombarci2007}, etc.
Lumps of SQM, the so-called strangelets \cite{Farhi1984},
or slets \cite{Peng2006PLB},
may exist in cosmic rays \cite{Shaulov2013},
and some of them might be on the way to our Earth \cite{Finch2006,Monreal2007}.
The neutron star could be converted to a quark star or mixed star
due to leptonic weak interactions, or seeded with slets by
the self-annihilating weakly interacting massive particles \cite{Perez-Garcia2010}.
The structure of a strange quark star depends strongly on the stability of
SQM which are still under active investigations \cite{Dexheimer2013}.
Presently, many aspects on quark matter are still left open.
Among them the equation of state (EOS) is of special interest.

In principle, one has in hand the fundamental theory of strong interactions,
i.e., quantum chromodynamics (QCD). Presently, however, no one can model
quark matter exactly in QCD because of the known difficulty
in the nonperturbative regime.
The only case that can be exactly solved is the noninteracting system
whose thermodynamic potential density
$ 
\Omega_0(T,\{\mu_i\},\{m_i\})
$ 
as a function of the temperature $T$, the chemical potentials $\{\mu_i\}$,
and the particle masses $\{m_i\}$
can be found in many textbooks.
For a free system, the particle mass is a constant, and the corresponding
thermodynamic treatment is clearly known.

Nowadays, quark matter has been investigated with
various phenomenological models,
e.g., the Richard potential model \cite{Bagchi2006,Sinha2013},
the Nambu and Jona-Lasinio (NJL) model \cite{Paoli2010},
the perturbation model \cite{Fraga2005},
the field correlator method \cite{Plumari2013},
the quark-cluster model \cite{XuRX2003},
and many other models~\cite{Khadekar2012,Sahoo2013,Isayev2013,Wang2010,%
WuC2005,Wen2013prd}, etc.
These models, to some extent, have a relation to, or start from, the free-particle system.
In the simplest version of the bag model \cite{Chodos1974,Witten1984,Farhi1984},
for example, a constant $B$, the so-called ``bag constant", is added to the
thermodynamic potential density of the free system to reflect
the quark confinement effect. This model has been applied
in a vast number of investigations on the properties of SQM
\cite{Madsen2000,Weber2010,Bialkowska2012,Rodrigues2013}.

It is well known, however, that particle masses vary with medium.
Such masses are usually called effective masses.
In principle, not only masses will change
but also the coupling constant will run in a medium \cite{Fraga2005}.
The models with chemical-potential and/or temperature dependent particle
masses are known as quasiparticle models \cite{Goloviznin1993},
which have been explored in great detail over the past two decades
\cite{Gorenstein1995,Schertler1997,Peshier2000,Bannur2007,Gardim2009}.
A recent example is the model without density or temperature
 dependent infinity of the vacuum zero-point energy \cite{ZongHS2012,ZongHS2013}.

Another important case is to include strong interquark
interactions with density and/or temperature dependent quark masses.
The original idea is to use a density dependent quark mass
to express nonperturbative interaction effects \cite{Fowler1981,Plumer1984}.
It was soon applied to study the equation of state (EOS)
\cite{Chakrabarty1989,Benvenuto1995,Peng2000,Yin2008,WangP2000,Peng1999,ZhangY2001,Wen2005},
the viscosity of SQM and dissipation of r-modes  \cite{Zheng2004},
the quark-diquark properties \cite{Lugones2003ijmpd},
and compact stars \cite{LiA2011}, etc.
Until now, these kinds of models have been developed greatly
\cite{Modarres2008,ChenSW2012,ChuPC2013,Torres2013}.

The most disputable issues in these kinds of models is the
thermodynamic inconsistency problem \cite{Chakrabarty1989,Benvenuto1995,Peng2000,Wen2005,Yin2008,WangP2000}.
Let us take the zero-temperature case to explain the issue.
Originally, all the thermodynamic formulas are taken as the same
of a free system \cite{Chakrabarty1989}. In this treatment (TD-1),
the properties of SQM are significantly different from
these in the conventional bag model. Subsequently, an additional term
was added to the pressure due to the density dependence of
quark masses, and simultaneously, the additional term was
subtracted from the energy density \cite{Benvenuto1995}.
In this second treatment (TD-2), SQM can be self-bound.
A serious problem is that the pressure
at the minimum of the energy per baryon
deviates obviously from zero.
The third treatment (TD-3) has the additional term
in the pressure to confine quarks,
but it does not appear in the energy density \cite{Peng2000}.
TD-3 successfully overcomes
the inconsistency between the zero pressure and energy minimum,
and later extended to finite temperature \cite{Wen2005}.

To study the deconfinement phase transition, one needs
to use true chemical potentials to maintain chemical equilibrium.
It was shown \cite{Peng2008} that the quark chemical
potentials used in the original TD-3 are in fact effective ones.
The true chemical potential $\mu_i$ of the quark flavor $i$
differs from the corresponding effective chemical potential $\mu_i^*$
by a common term for all quark flavors,
and accordingly satisfies the same weak equilibrium conditions.
For SQM, therefore, the effective chemical
potentials act like the real chemical potentials to give the same EOS.

Recently, another effort has been made
to clear the ambiguity in thermodynamic treatments, where
the quark mass was regarded as an intrinsic freedom, and
an additional term was added to the fundamental thermodynamic
differential equation \cite{Yin2008}.
Assuming the effective mass intrinsic while it depends completely
on the state variables (the density and/or temperature) is
conceptually self-contradictory, and inevitably leads to inconsistency.
In fact, with the additional term to the fundamental thermodynamic
differential equation, the original TD-1 treatment was recovered.
Because the pressure in this treatment is always positive,
which poses a problem related to the stability of SQM,
the authors finally had to consider the vacuum contribution
by adding a term to the thermodynamic potential density,
as had been done in a previous reference~\cite{WangP2000},
to which we refer as TD-4.

Another important aspect very relevant to thermodynamic treatment
is how the quark masses depend on the density and/or temperature.
Originally, the density dependence of quark masses is parametrized as
\begin{equation} \label{linsca}
m_i=m_{i0}+\frac{B}{3n_\mathrm{b}},
\end{equation}
which was first given for light quarks in Ref.~\cite{Fowler1981}
according to bag model assumption,
and extended to including strange quarks in Ref.~\cite{Chakrabarty1989}.
An alternative parametrization of the density dependence is the cubic-root scaling
\begin{equation} \label{cubicsca}
m_i=m_{i0}+\frac{D}{n_\mathrm{b}^{1/3}},
\end{equation}
derived from the linear confinement and
 leading-order in-medium chiral condensate \cite{Peng1999}.
In Eqs. (\ref{linsca}) and (\ref{cubicsca}),
$m_{i0}$ ($i=u,d,s$) are the corresponding quark current mass,
$n_\mathrm{b}$ is the baryon number density,
$B$ and $D$ are constants signifying the confinement strength.

The purposes of the present paper are twofold.
First, in the next section, we give the quantitative
 criteria for thermodynamic consistency,
and compare the above-mentioned treatments. It is explicitly shown that
TD-1 and TD-2 have unreasonable vacuum limits, and their pressure
at the minimum of energy (free energy at finite temperature) per baryon
deviates obviously from zero, contradicting standard thermodynamics.
The existence of an additional term to the thermodynamic potential density in TD-4 depends
on whether the quark mass scaling satisfies the Cauchy condition
that ensures the integrability of the relevant path integral.
Unfortunately, however, neither Eq.~(\ref{linsca}) nor Eq.~(\ref{cubicsca})
meets the requirement. Therefore, the added term in TD-4 for thermodynamic consistency
does not exist for the presently known quark mass scaling, and consequently
TD-4 itself violates the Maxwell integrability condition.
On the other hand, TD-3 with effective chemical potential interpretation
can naturally give consistent vacuum limits, and the pressure at
the minimum energy (free energy at finite temperature) is exactly zero.
At the same time, all expressions in TD-3 are given explicitly without
any integral, and more importantly, the Maxwell condition are fulfilled.

Recently, there is much progress in the measurement of compact stars with mass
about 2 times the solar mass (2$M_\odot$) \cite{Antoniadis2013,Demorest2010}.
It is shown that
the reciprocity scaling in Eq.~(\ref{linsca}) together with TD-2
generates more massive strange stars than in the bag model
\cite{Torres2013}.
However, the thermodynamic treatment, TD-2 used there,
suffers from thermodynamic inconsistency, as the authors also noticed.
On the other hand, the thermodynamically consistent TD-3 with
the cubic-root scaling can describe
stars with radii even smaller than in the bag model \cite{Dexheimer2013},
but the maximum mass was normally much smaller than $2M_\odot$ \cite{Peng2000}.
This case is mainly because the presently known quark mass scaling,
either the reciprocity scaling or the cubic-root scaling,
merely takes account of the confinement interaction, while
the important perturbative interactions are unable to be included.

It is, therefore, our second purpose in the present paper
to look for a new
quark mass scaling which considers both the confinement and leading-order
perturbative interactions. We show that this new scaling with TD-3 can describe
quark stars with both low and high maximum masses, depending on the confinement
and perturbative strength parameters. If SQM is absolutely stable,
the maximum mass is as large as 2 times the solar mass, consistent with
the recent measurements \cite{Antoniadis2013,Demorest2010}.

In Sec.~\ref{inconThermo},
we first compare the four thermodynamic treatments.
 Then in Sec.~\ref{sec:therm},
we present a fully consistent derivation of the thermodynamic treatment
that will be used in the present paper.
 After arriving at a new quark mass scaling
which includes both confinement and leading-order perturbative interactions
in Sec.~\ref{sec:mass}, the properties of SQM and
the mass radius relation of strange stars are
calculated in the new scaling in Secs.~\ref{sec:SQM} and ~\ref{stars}.
A summary is given in the final Sec.~\ref{sec:sum}.

\section{Inconsistency of the recent thermodynamic treatment}
\label{inconThermo}

For comparison purpose, let us start from the fundamental
differentiation equation of standard thermodynamics,
\begin{equation} \label{dEbar}
\mbox{d}\bar{E}=T\mbox{d}\bar{S}-P\mbox{d}V+\sum_i\mu_i\mbox{d}\bar{N}_i,
\end{equation}
where $\bar{E}$ is the internal energy, $T$ the temperature, $\bar{S}$ the entropy,
$\bar{N}_i$ the particle number of particle type $i$,
and $\mu_i$ the corresponding chemical potential.
The three terms on the right are, respectively, the three
ways of increasing the system internal energy, i.e.,
heat transfer, doing work, and particle exchange.

Defining the free energy
\begin{equation} \label{Fbardef}
\bar{F}\equiv\bar{E}-T\bar{S},
\end{equation}
Eq.~(\ref{dEbar}) then becomes
\begin{equation} \label{dFbar}
\mbox{d} \bar{F}
=-\bar{S}\mbox{d}T -P\mbox{d}V +\sum_i\mu_i\mbox{d}\bar{N}_i.
\end{equation}

Similarly using the thermodynamic potential,
\begin{equation} \label{Omegabardef}
\bar{\Omega}
\equiv \bar{F}-\sum_i\mu_i\bar{N}_i
=\bar{E}-T\bar{S}-\sum_i\mu_i\bar{N}_i,
\end{equation}
then Eq.~(\ref{dEbar}) or Eq.~(\ref{dFbar}) becomes
\begin{equation} \label{dOmegabar}
\mbox{d}\bar{\Omega}
=-\bar{S}\mbox{d}T-P\mbox{d}V-\sum_i\bar{N}_i\mbox{d}\mu_i.
\end{equation}

In Ref.~\cite{Yin2008}, the quark mass was assumed to be
an intrinsic degree of freedom, and Eq.~(\ref{dOmegabar}) was modified to
\begin{equation} \label{dOmegabarYin}
\mbox{d}\bar{\Omega}
=-\bar{S}\mbox{d}T-P\mbox{d}V-\bar{N}\mbox{d}\mu+X\mbox{d}m^*,
\end{equation}
when the system has only one type of particles and
the particle mass is density and/or temperature dependent, i.e.,
\begin{equation}
m^*=m^*(T,n_\mathrm{b}).\
\end{equation}

Equations~(\ref{dOmegabar}) and (\ref{dOmegabarYin}) have
significantly different thermodynamic consequences.
In the latter case, all the thermodynamic formulas
have the same form as these of a free system,
as in Eq. (45) of Ref.~\cite{Yin2008}, i.e.,
$$
\begin{array}{rll}
& \bar{S}=-\left(\frac{\partial\bar{\Omega}}{\partial T}\right)_{V,\mu,m}, \ \
& P=-\left(\frac{\partial\bar{\Omega}}{\partial V}\right)_{T,\mu,m}, \\
&\bar{N}
=-\left(\frac{\partial\bar{\Omega}}{\partial\mu}\right)_{T,V,m}, \ \
&X=\left(\frac{\partial\bar{\Omega}}{\partial m}\right)_{T,V,\mu}.
\end{array}
$$

%

In physics, the effective mass is a phenomenological concept.
It is usually introduced to meet some special purposes.
For example, in the relativistic mean-field theory,
an effective nucleon mass,
$M^*=M_\mathrm{N}-g\phi$
with $g$ the coupling and $\phi$\ the scalar field,
was introduced to make the equation of motion become the Dirac equation
\cite{Walecka1995}; in the NJL model, the quark's effective mass
was introduced as $m_q^*=M_q-G\langle\bar{q}q\rangle$\ \cite{Henley1990};
to avoid complexity of nonlocality in the many-body theory of the early stage,
the interacting nucleon was treated as a particle with
kinetic energy $E_k=p^2/(2M^*)+a$ where the effective mass
satisfies $1/M^*=1/M+2b$ \cite{Marmier1970};
in quantum hadrondynamics,
an effective mass of the nucleon was determined by
a self-consistent equation \cite{Serot1997}; etc.

In the present case, the density and/or temperature dependent quark mass
is also conceptually effective.
It is logically contradictory to assume the
quark mass intrinsic while it is, in fact, dependent on,
or determined by, the medium state variables (the density and/or temperature).
The thermodynamic treatment derived from
Eq.~(\ref{dOmegabarYin}) is thus inevitably inconsistent with the standard thermodynamics.

\begin{figure}
\includegraphics[width=8cm]{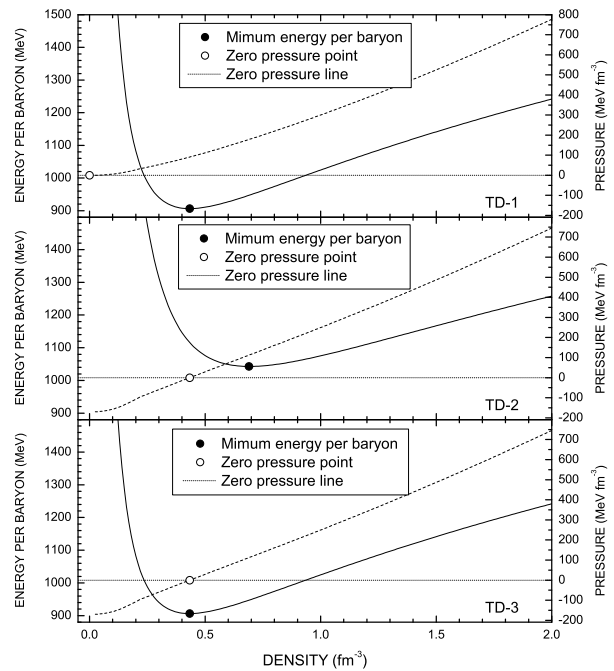}
\caption{Comparison of the thermodynamic treatments
TD-1 (upper panel), TD-2 (middle panel), and TD-3 (lowest panel).
The quark mass scaling parameters are the same as in Ref.~\cite{Yin2008}
for Figs.~1 and 2 there. The energy per baryon and pressure are,
respectively, on the left and right axis.
The pressure is obviously nonzero at the minimum energy per baryon
for TD-1 and TD-2, and exactly zero for TD-3.
\label{figYin}}
\end{figure}

To explicitly show the inconsistency, we have proved
a necessary condition in the Appendix.
At zero temperature, the condition is
\begin{equation} \label{DeltadefT0}
\Delta
=P-n_\mathrm{b}^2
 \frac{\mathrm{d}}{\mathrm{d}n_\mathrm{b}}
 \left(
  \frac{E}{n_\mathrm{b}}
 \right)=0,
\end{equation}
where $P$ and $E$ are the model-given thermodynamic expressions.
Any consistent thermodynamic treatment must ensure
that $\Delta$\ be zero at arbitrary density. Specifically,
the pressure at the minimum energy per baryon must be zero.

\begin{table} \label{EminPzero}
\caption{
Pressure at the minimum of energy per baryon calculated by the
different thermodynamic treatment approaches
TD-1 \cite{Yin2008,Chakrabarty1989}, TD-2 \cite{Benvenuto1995},
and TD-3.
        }
\begin{tabular}{cccccccccc} \hline
  & ${\cal E}_{\mathrm{min}}$ & & Density && Pressure && $n_0$ && ${\cal E}_0$  \\
  &  (MeV) & & (fm$^{-3}$) && (MeV fm$^{-3}$)&& (fm$^{-3}$) &&  (MeV)
               \\ \cline{2-2}\cline{4-4}\cline{6-6}\cline{8-8}\cline{10-10}
TD-1 &  906.3 && 0.433 && 90.7  && 0 && $\infty$ \\
TD-2 & 1041.8 && 0.692 && 113.2 && 0.433 && 1116.2\\
TD-3 &  906.3 && 0.433 && 0     && 0.433 && 906.3 \\ \hline
\end{tabular}
\end{table}

In the upper panel of Fig.~\ref{figYin}, we reproduce the Fig.~1 of Ref.~\cite{Yin2008},
with the pressure added on the right axis. The cases for the second
and third treatments are also given, respectively,
in the middle and lowest panels for comparison.
The minimum energy per baryon (${\cal E}_\mathrm{min}$),
its position density, the pressure at the minimum,
the position of zero pressure ($n_0$),
and the energy per baryon at the zero pressure (${\cal E}_0$)
are listed in Table 1.
It is obviously seen that the pressure at the energy minimum is nonzero
for TD-1 and TD-2, and they are exactly located at the same density for TD-3.

\begin{figure}
\centering
\includegraphics[width=8cm]{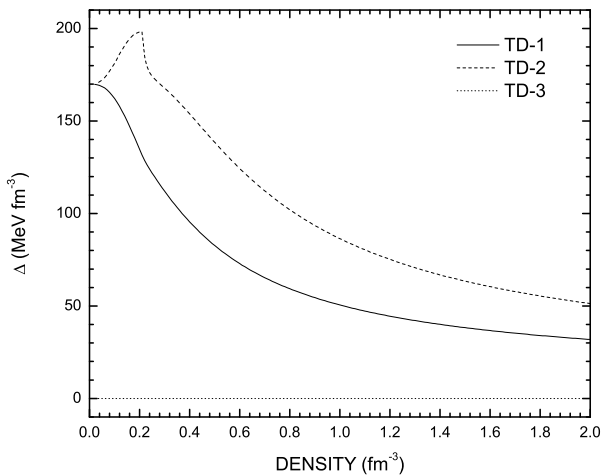}
\caption{
The $\Delta$ as a function of density at zero temperature.
The solid, dashed, and dotted lines are, respectively
for the thermodynamic treatments TD-1, TD-2, and TD-3.
\label{figDelta}
        }
\end{figure}

In Fig.~\ref{figDelta}, we give the $\Delta$\ as a function of density
with the same mass scaling and parameters
[Eq.~(\ref{linsca}) with
$
m_{u0}=m_{d0}=0, m_{s0}=150\ \mbox{MeV},
B^{1/4}=170\ \mbox{MeV fm}^{-3}].
$
Again, we see that both TD-1 and TD-2 cannot give zero $\Delta$,
while TD-3 naturally gives $\Delta=0$ at the whole density region.

Aside from the nonzero $\Delta$ at arbitrary density
and the nonzero pressure at the energy minimum,
the vacua in TD-1 and TD-2 are also contradictory themselves.
Using the reciprocity scaling in Eq.~(\ref{linsca}), one can easily obtain
\begin{eqnarray}
&&\lim_{n_\mathrm{b}\rightarrow 0} E_1=B, \
  \lim_{n_\mathrm{b}\rightarrow 0} P_1=0.
 \label{EPi01}  \\
&&\lim_{n_\mathrm{b}\rightarrow 0} E_2=2B, \
  \lim_{n_\mathrm{b}\rightarrow 0} P_2=-B.
 \label{EPi02} \\
&&\lim_{n_\mathrm{b}\rightarrow 0} E_3=B, \
  \lim_{n_\mathrm{b}\rightarrow 0} P_3=-B,
\end{eqnarray}
where $E_i$ and $P_i$ ($i=1, 2, 3$)
are the corresponding energy density and pressure
in the three thermodynamic treatments.
The vacua in TD-1 and TD-2 violate
the universal energy conservation law.

Therefore, modifying the fundamental thermodynamic equation
causes unavoidable inconsistencies.
In fact, one should not modify the fundamental equation (\ref{dOmegabar}).
Instead, we should modify phenomenological models to meet its requirements.
In the next section, we will explain such a scheme to make
the previous thermodynamic treatment in full consistency with
the standard thermodynamics.

Because the pressure in TD-1 cannot be negative,
another treatment was recommended in Sec.~V of Ref.~\cite{Yin2008}.
The recommended treatment follows that in Ref.~\cite{WangP2000},
where the total thermodynamic density was written as
\begin{equation}
\Omega
= \Omega_0(T,\{\mu_i\},\{m_i\})
 +\Omega_\mathrm{a}(n_{\mathrm{b}}).
\end{equation}
The functional form of $\Omega_0$ is the same as a free-particle system.

Adding a new term to the thermodynamic potential density
is a good idea because the thermodynamic potential density
changes when interactions set in.
In the quasiparticle model, the added term
can be self-consistently obtained when the particle mass
depends solely on temperature \cite{Gorenstein1995},
or on respective chemical potentials [$m_i^*=m_i^*(\mu_i)$ when the chemical
potentials are not coupled] \cite{Schertler1997},
or even the finite size effects are considered \cite{Wen2010}.

In the case of density and/or temperature dependent masses, however,
chemical potentials are surely coupled. We therefore
need to check if the added term really exists for the presently
known quark mass scaling.

Let's first see how the added term is determined
in the literature \cite{Yin2008,WangP2000}.
Similarly as in the quasiparticle model, the extra term
was chosen to be determined by
\begin{equation} \label{dOmegaWP}
\frac{\mathrm{d}\Omega_{\mathrm{a}}}{\mathrm{d}n_{\mathrm{b}}}
=-\sum_i\frac{\partial\Omega_0}{\partial m_i}
  \frac{\mathrm{d}m_i}{\mathrm{d}n_{\mathrm{b}}}
\end{equation}
so that the particle number density was still given by
\begin{equation} \label{niori}
n_i
=-\frac{\partial\Omega_0}{\partial\mu_i}
=\frac{g_i}{6\pi^2}\left(\mu_i^2-m_i^2\right)^{3/2}.
\end{equation}
Then from Eq.~(\ref{dOmegaWP}) the added term was obtained as
\begin{equation} \label{OmegaWP}
\Omega_\mathrm{a}(n_{\mathrm{b}})
=-\int_{\rho_c}^{n_{\mathrm{b}}} \sum_i\frac{\partial\Omega_0}{\partial m_i}
  \frac{\mathrm{d}m_i}{\mathrm{d}n_{\mathrm{b}}} \mbox{d}n_{\mathrm{b}}
  +\Omega_a(\rho_c),
\end{equation}
where $\rho_c$ is an integral constant.

In the integrand of Eq.~(\ref{OmegaWP}), there is not only
the argument $n_{\mathrm{b}}$, but also the chemical
potentials $\mu_i$ ($i=u,d,s$). Because these chemical potentials
are not constants [otherwise one can inverse Eq.~(\ref{OmegaWP})
to go back to the bag model],
they must be determined as functions of the density:
\begin{equation} \label{SpePath}
\mu_i=\mu_i(n_{\mathrm{b}}).
\end{equation}

In Refs.~\cite{WangP2000,Yin2008},
the functions $\mu_i(n_{\mathrm{b}})$ were chosen
by solving Eqs.
(\ref{weakequi})-(\ref{Baryonconserv}).
Because the thermodynamic potential density
is a state quantity, however, the integration should be independent
of the special path. In the following, we show that
only when the Cauchy condition is satisfied, can one obtain the additional term
by choosing a special path.

In fact, to eliminate the extra term in the expression of particle numbers due to the density dependence of quark masses, one must require
\begin{equation} \label{dOmegaa}
\frac{\partial\Omega_a}{\partial\mu_i}
=-\sum_k\frac{\partial\Omega_0}{\partial m_k}
  \frac{\mathrm{d}m_k}{\mathrm{d}n_{\mathrm{b}}}
  \frac{\partial n_{\mathrm{b}}}{\partial \mu_i}.
\end{equation}
Namely, the additional term should be given by a path integral as
\begin{eqnarray}
\Omega_{\mathrm{a}}
&=&
  -\int_{\mu_0}^{\mu}\sum_i
  \frac{\partial\Omega_a}{\partial\mu_i}
  \mbox{d}\mu_i
\nonumber\\
&=&
-\int_{\mu_0}^{\mu}
  \left(
  \sum_k\frac{\partial\Omega_0}{\partial m_k}
  \frac{\mathrm{d}m_k}{\mathrm{d}n_{\mathrm{b}}}
  \right)
  \sum_i \frac{\partial n_{\mathrm{b}}}{\partial \mu_i} \mbox{d}\mu_i.
\label{Omegapath}
\end{eqnarray}

When one chooses a special path such as that in Eq.~(\ref{SpePath}),
Eq.~(\ref{Omegapath}) leads to Eq.~(\ref{OmegaWP}).
Because $\Omega_{\mathrm{a}}$ is a state function of the independent
state quantities $\mu_i$ ($i=u, d, \mbox{and}\ s$), the path integral
on the right-hand side of Eq.~(\ref{Omegapath}) should be path independent.
 For this the famous Cauchy theorem must be satisfied:
\begin{equation}
\frac{\partial^2\Omega_{\mathrm{a}}}
     {\partial\mu_j\partial\mu_i}
=
\frac{\partial^2\Omega_{\mathrm{a}}}
     {\partial\mu_i\partial\mu_j}.
\end{equation}
Substituting the right-hand side of Eq.~(\ref{dOmegaa}) then gives
\begin{equation}
\frac{
\sum_k \frac{\partial^2\Omega_0}{\partial\mu_i\partial m_k}
       \frac{\mathrm{d}m_k}{\mathrm{d}n_{\mathrm{b}}}
     }
     {
 \partial n_{\mathrm{b}}/\partial\mu_i
     }
 =
\frac{
\sum_k \frac{\partial^2\Omega_0}{\partial\mu_j\partial m_k}
       \frac{\mathrm{d}m_k}{\mathrm{d}n_{\mathrm{b}}}
     }
     {
 \partial n_{\mathrm{b}}/\partial\mu_j
     }
\end{equation}

On application of Eq.~(\ref{niori}), one has
\begin{equation}
\frac{\partial n_{\mathrm{b}}}{\partial\mu_i}
=\frac{\mu_i\sqrt{\mu_i^2-m_i^2}}
      {\frac{6\pi^2}{g} + \sum_k m_k \sqrt{\mu_k^2-m_k^2} \frac{\mathrm{d}m_k}{\mathrm{d}n_{\mathrm{b}}} }.
\end{equation}
Then one can obtain
\begin{equation} \label{CaushyCond}
 \frac{m_u}{\mu_u} \frac{\mathrm{d}m_u}{\mathrm{d}n_{\mathrm{b}}}
=\frac{m_d}{\mu_d} \frac{\mathrm{d}m_d}{\mathrm{d}n_{\mathrm{b}}}
=\frac{m_s}{\mu_s} \frac{\mathrm{d}m_s}{\mathrm{d}n_{\mathrm{b}}}.
\end{equation}

Unfortunately, however, neither the reciprocity scaling \cite{Fowler1981}
nor the cubic-root scaling satisfies the Cauchy condition (\ref{CaushyCond}).
An even more general form of $m_i=m_{i0}+m_\mathrm{I}(n_\mathrm{b})$
cannot do so. Therefore, the additional term
$\Omega_{\mathrm{a}}$ does not exist for the
presently known density and/or temperature dependent quark mass scaling.
We should therefore look for another more convenient treatment
that is in agreement with the fundamental Eq.~(\ref{dOmegabar}),
or equivalently, Eq.~(\ref{dFbar}).

\section{self-consistent thermodynamics
with density and/or temperature dependent particle masses}
\label{sec:therm}


Because the system we are studying has density and/or temperature
dependent masses, it is naturally convenient to choose
the temperature $T$, the densities $n_i$,
and the volume $V$ as the independent state variables.
Therefore, we rewrite the
fundamental Eq.~(\ref{dFbar}) as
\begin{equation}
\mbox{d}F
=-S\mbox{d}T +\left(-P-F+\sum_i\mu_in_i\right)\frac{\mbox{d}V}{V} + \sum_i\mu_i\mbox{d}n_i
\end{equation}
where $F\equiv \bar{F}/V,\ S\equiv\bar{S}/V,\ n_i\equiv\bar{N}_i/V$
are, respectively, the free-energy density, the entropy density,
and the particle number densities.

For an infinitely large system such as quark matter,
the free-energy density has nothing to do with the volume.
In this case, we have
\begin{equation} \label{Pinf}
P=-F+\sum_i\mu_in_i
\end{equation}
and
\begin{equation} \label{dFinf}
\mbox{d}F
=-S\mbox{d}T +\sum_i\mu_i\mbox{d}n_i.
\end{equation}

Now we try to establish a thermodynamic treatment
in full agreement with this fundamental equation.
Because both TD-1 and TD-2 do not give zero pressure
at the energy minimum, they should naturally be discarded.
The model we are trying to build resembles, in many aspects,
the third treatment TD-3.

At zero temperature, the energy density of a free quark system
is
\begin{equation} \label{E0nu}
E_0=\sum_ig_i\int_0^{\nu_i}
   \sqrt{p^2+m_i^2}\frac{p^2\mathrm{d}p}{2\pi^2},
\end{equation}
where $\nu_i$\ is the particle type $i$'s Fermi momentum
which is connected to the corresponding particle number density by
\begin{equation}
n_i=g_i\int_0^{\nu_i}\frac{p^2\mathrm{d}p}{2\pi^2}
=\frac{g_i\nu_i^3}{6\pi^2}.
\end{equation}
When the quarks interact with each other, we want to include the
interaction effect with a density dependent mass as
\begin{equation}
m_i=m_{i0}+m_\mathrm{I},
\end{equation}
where $m_\mathrm{I}$ is a density dependent quantity.
We demand that the system energy density still has the same form
as in Eq.~(\ref{E0nu}), i.e., $E=E_0$, which is also the original idea
in Ref.~\cite{Fowler1981}. This is possible
because $E_0$ is an increasing function of the particle masses.
To distinguish with other mass concepts, we call such a
mass an equivalent mass \cite{Peng2003}.
With the equivalent mass,
both the energy density and particle number densities have the same form as
a free particle system while only the particle number densities keep unchanged in
the quasiparticle model. We call such a model an equiparticle model.
The corresponding pressure in the equiparticle model can then be easily deduced
according to the fundamental thermodynamics, as in
Sec. II of Ref.~\cite{Peng2008}. Here we do not repeat the derivation.
The key point is that the Fermi momentum $\nu_i$
is not directly linked to the chemical potential $\mu_i$ by
$\nu_i=\sqrt{\mu_i^2-m_i^2}$; instead, it is connected to an effective chemical
potential $\mu_i^*$ by $\nu_i=\sqrt{{\mu_i^*}^2-m_i^2}$, while
the relation between the effective and real chemical potentials
are determined by the fundamental differential equation (\ref{dFinf}),
and, consequently, the pressure, as well as the thermodynamic potential density
have an additional term due to the density dependence  of quark masses.

At finite temperature, the concept of the Fermi momentum
is not as useful as in the case of zero temperature.
We should directly use the concept of effective chemical potentials.
Also the characteristic function should be changed from the energy
to the free energy.
Therefore, we write the free-energy density of the system
the same as that of a free system with equivalent particle mass $m_i(T,n_\mathrm{b})$
and effective chemical potentials $\mu_i^*$ at temperature $T$, i.e.,
\begin{equation} \label{Fmod}
F=\Omega_0(T,\{\mu_i^*\},\{m_i\})+\sum_i\mu_i^*n_i.
\end{equation}
Please note the arguments in $\Omega_0$: the position
of free particle's chemical potentials have been replaced with the
effective chemical potentials $\mu_i^*$, i.e.,
\begin{equation}
\Omega_0=\Omega_0(T,\{\mu_i^*\},\{m_i\})
\end{equation}
is the thermodynamic potential density of a free system
with the particle masses $m_i(T,n_\mathrm{b})$ and chemical potentials $\mu_i^*$.
Because the independent state variables are ($T,V,\{n_i\}$),
not including $\mu_i^*$, we should also choose how to connect
$\mu_i^*$ to the independent variables. Here we choose
to connect the effective chemical potentials to particle
number densities by
\begin{equation} \label{nimod}
n_i=-\frac{\partial}{\partial\mu_i^*}\Omega_0(T,\{\mu_i^*\},\{m_i\}),
\end{equation}
which are also the choice of many previous thermodynamic treatments,
but the real chemical potentials have been replaced with effective ones here
to ensure the thermodynamic consistency.

To derive other thermodynamic quantities,
let us differentiate Eq.~(\ref{Fmod}) to give
\begin{equation} \label{dFtmp}
\mbox{d}F
=\mbox{d}\Omega_0 +\sum_i n_i\mbox{d}\mu_i^* +\sum_i\mu_i^*\mbox{d}n_i,
\end{equation}
where
\begin{equation} \label{dOmega0gen}
\mbox{d}\Omega_0
= \frac{\partial\Omega_0}{\partial T}\mbox{d}T
 +\sum_i\frac{\partial\Omega_0}{\partial\mu_i^*}\mbox{d}\mu_i^*
 +\sum_i\frac{\partial\Omega_0}{\partial m_i}\mbox{d}m_i
\end{equation}
with
\begin{equation} \label{dmigen}
\mbox{d}m_i
= \frac{\partial m_i}{\partial T}\mbox{d}T
 +\sum_j\frac{\partial m_i}{\partial n_j} \mbox{d}n_j.
\end{equation}

On application of Eqs.~(\ref{nimod}), (\ref{dOmega0gen}), and (\ref{dmigen}),
Eq.~(\ref{dFtmp}) becomes
\begin{eqnarray}
\mbox{d}F
&=&
 \left(
   \frac{\partial\Omega_0}{\partial T}
  +\sum_i\frac{\partial\Omega_0}{\partial m_i}\frac{\partial m_i}{\partial T}
 \right)\mbox{d}T
\nonumber\\
& &
 +\sum_i\left(
    \mu_i^*
   +\sum_j\frac{\partial\Omega_0}{\partial m_j}\frac{\partial m_j}{\partial n_i}
  \right)\mbox{d}n_i.
\end{eqnarray}
Comparing this equation with Eq.~(\ref{dFinf}), we immediately have
the entropy density
\begin{equation} \label{Smod}
S=-\frac{\partial\Omega_0}{\partial T}
  -\sum_i\frac{\partial\Omega_0}{\partial m_i}\frac{\partial m_i}{\partial T}
\end{equation}
and the true chemical potential
\begin{equation} \label{mumustarT}
\mu_i
= \mu_i^*
 +\sum_j\frac{\partial\Omega_0}{\partial m_j}\frac{\partial m_j}{\partial n_i}
\equiv \mu_i^*-\mu_\mathrm{I}.
\end{equation}

The pressure can be obtained by substituting Eq.~(\ref{Fmod})
into Eq.~(\ref{Pinf}), giving
$P=-\Omega_0+\sum_i(\mu_i-\mu_i^*)n_i$, i.e.,
\begin{equation} \label{Pmod}
P=-\Omega_0
  +\sum_{i,j}\frac{\partial\Omega_0}{\partial m_j}
             n_i\frac{\partial m_j}{\partial n_i}.
\end{equation}
The energy density is obtained by substituting
Eqs.~(\ref{Fmod}) and (\ref{Smod}) into $E=F+TS$ as
\begin{equation}
E= \Omega_0
 -\sum_i\mu_i^*\frac{\partial\Omega_0}{\partial\mu_i^*}
 -T\frac{\partial\Omega_0}{\partial T}
 -T\sum_i\frac{\partial\Omega_0}{\partial m_i}
         \frac{\partial m_i}{\partial T},
\end{equation}
while the real thermodynamic potential density is
\begin{equation} \label{Omegamod}
\Omega
=F-\sum_i\mu_in_i
=\Omega_0
  -\sum_{i,j}\frac{\partial\Omega_0}{\partial m_j}
             n_i\frac{\partial m_j}{\partial n_i}.
\end{equation}

For a given set of the independent state variables $T$ and $n_i$,
the effective chemical potential $\mu_i^*$ is obtained by solving
the equation(s) in Eq.~(\ref{nimod}). Then other thermodynamic quantities
can be calculated, respectively, by Eqs.~(\ref{Smod})-(\ref{Omegamod})
if the temperature and density dependence of the quark masses is known.

From Eqs.~(\ref{Pmod}) and (\ref{Omegamod}), one finds that the normal relation
$P=-\Omega$ still holds.
In fact, all the basic relations of standard thermodynamics
are maintained in the present treatment.
The $\Omega_0$, as seen from the derivation process,
serves merely as an intermediate quantity,
while other thermodynamic quantities are derived and expressed
in its functional form.
In the following, for example,
we list formulas for the two specially important cases at zero temperature.

\subsection{Color-flavor locking with density-dependent
particle masses}

The color-flavor locked phase is believed to exist at
extremely high density \cite{Rajagopal2000,Alford2001}.
In the MIT bag model, one has known how to construct the thermodynamic
density long ago \cite{Alford2001prd,Lugones2002,Peng2006}.
To consider the medium effect with density-dependent quark masses,
one can similarly construct the thermodynamics density \cite{Chen2007,Wen2007}.

The free particle contribution is
\begin{eqnarray}
\Omega_0
&=&
 \sum_q\frac{3}{\pi^2}
  \int_0^{\nu}
  \left(\sqrt{p^2+m_q^2}-\mu_q^*\right)p^2\mbox{d}p
\nonumber\\
&&
 +\frac{3\Delta^2}{\pi^2}\bar{\mu}^2 +B.
\end{eqnarray}
where the chemical potentials have been replaced with the
effective ones to consider medium effect with density dependent
particle masses, the second term is the paring contribution
with $\bar{\mu}=(\mu_u^*+\mu_d^*+\mu_s^*)/3$,
and the last term is added if one would also like to
include the vacuum contribution. The common Fermi momentum
in the first term is obtained by minimizing $\Omega_0$
at fixed $\mu_q^*$, i.e.,
$
{\partial\Omega_0}/{\partial\nu}
=0\
$
which gives
\begin{equation}
\sum_q\sqrt{\nu^2+m_q^2}=3\bar{\mu}.
\end{equation}

All other thermodynamic quantities can now be directly obtained
from the above formulas. They are the same as
in Ref.~\cite{Chen2007},
and including finite-size effects in Ref.~\cite{Wen2007},
with the emphasis that the chemical potentials there
be regarded as effective ones according to
the above consistent thermodynamic derivations.

\subsection{The unpaired case at zero temperature}

For the unpaired SQM at finite temperature, we have the same formulas
as in Ref.~\cite{Peng2008}, or including finite-size
effects in Ref.~\cite{Wen2005}.
At zero temperature, the formulas are still the same as
those in Ref.~\cite{Peng2000}.
Again, the chemical
potentials there should be regarded as effective ones.

For the convenience of getting a new quark mass scaling
in the next section, we give the free unpaired particle contribution:
\begin{equation} \label{Omega0T0}
\Omega_0
=-\sum_i\frac{g_i}{24\pi^2}
 \left[ \mu_i^*\nu_i(\nu_i^2-\frac{3}{2}m_i^2)
 +\frac{3}{2} m_i^4\ln\frac{\mu_i^*+\nu_i}{m_i}
 \right].
\end{equation}
Correspondingly, we have the particle number density
\begin{equation} \label{niT0}
n_i
=\frac{g_i}{6\pi^2}
 \left(
  {\mu_i^*}^2-m_i^2
 \right)^{3/2}
=\frac{g_i\nu_i^3}{6\pi^2},
\end{equation}
the relation between the real and effective chemical potentials
\begin{equation} \label{mumustar}
\mu_i
= \mu_i^*
 +\frac{1}{3}\frac{\partial m_{\mathrm{I}}}{\partial n_{\mathrm{b}}}
             \frac{\partial\Omega_0}{\partial m_{\mathrm{I}}}
\equiv \mu_i^*-\mu_{\mathrm{I}},
\end{equation}
and the pressure
\begin{equation} \label{PT0}
P=-\Omega_0
  +n_{\mathrm{b}}
   \frac{\partial m_{\mathrm{I}}}{\partial n_{\mathrm{b}}}
   \frac{\partial\Omega_0}{\partial m_{\mathrm{I}}}
 =-\Omega_0-3n_\mathrm{b}\mu_\mathrm{I}.
\end{equation}

In Eqs.~(\ref{Omega0T0})-(\ref{PT0}),
\begin{equation}
\nu_i=\sqrt{{\mu_i^*}^2-m_i^2}
\end{equation}
is the Fermi momentum of particle type $i$,
while the derivative of $\Omega_0$ with respect to the interacting quark mass
is
\begin{equation}
\frac{\partial\Omega_0}{\partial m_{\mathrm{I}}}
=\sum_i \frac{g_im_i}{4\pi^2}
 \Bigg[
  \mu_i^*\nu_i 
  -m_i^2\ln\frac{\mu_i^*+\nu_i}{m_i}
 \Bigg].
\end{equation}

For future convenience, we define a holistic Fermi momentum as
\begin{equation} \label{nudefn}
\nu \equiv
\left(
 \frac{1}{N_\mathrm{f}} \sum_{q}\nu_q^3
\right)^{1/3},
\end{equation}
where the summation index $q$ goes over all quark flavors involved,
and $N_\mathrm{f}$ is the quark flavor number.
Equation~(\ref{nudefn}) means $\nu$\ is the
subtriplicate of the averaged cubic Fermi momentum.
With a view to Eq.~(\ref{niT0}) and the definition of the baryon number density
$n_\mathrm{b}=\sum_qn_q/3$,
Eq.~(\ref{nudefn}) can naturally be linked to density by
\begin{equation} \label{nunb}
\nu
=\left(
  \frac{3\pi^2}{N_\mathrm{f}} n_\mathrm{b}
 \right)^{1/3},\ \ \
n_\mathrm{b}=\frac{N_\mathrm{f}}{3\pi^2}\nu^3.
\end{equation}
Obviously, $\nu$\ has the dimension of energy. It can therefore be used as
an energy scale of cold quark matter. For the color-flavor locked case, it equals to the common fermi momentum.

\section{Quark mass scaling with linear confinement and leading-order perturbative interactions}
\label{sec:mass}

In order to get an appropriate equivalent mass, we carry out the similar procedure as was done in Ref~\cite{Peng2005npa},
namely we expand the equivalent mass to a Laurant series
of the holistic Fermi momentum $\nu$, and take the leading term in both directions:
\begin{equation} \label{mInu}
m_\mathrm{I}=\frac{a_{-1}}{\nu} + a_1\nu.
\end{equation}

We will soon see that the first term corresponds to
the linear confinement, while the second term is responsible for
the leading-order perturbative interactions.

At lower density, the first term becomes infinitely large when
the holistic Fermi momentum, or the density, approaches to zero.
Therefore, Eq.~(\ref{mInu}) becomes $m_\mathrm{I}=a_{-1}/\nu$ at lower density.
On the other hand, we have already known that the lower density behavior is
$m_\mathrm{I}=D/n_{\mathrm{b}}^{1/3}$ with
the confinement parameter $D$ connected to
the string tension $\sigma_0$, the chiral restoration density $\rho^*$,
and the sum of the vacuum chiral condensates $\sum_q\langle\bar{q}q\rangle_0$ by
\begin{equation}
D\sim\frac{3(2/\pi)^{1/3}\sigma_0\rho^*}
{-\sum_q\langle\bar{q}q\rangle_0}.
\end{equation}
Although we cannot use this formula to exactly calculate the $D$
value due to the uncertainties in relevant quantities,
we do know that $D$ is a low energy parameter, and that $\sqrt{D}$ approximately
is in the range of (147, 270) MeV \cite{Wen2005}.
Equating $a_{-1}/\nu$ and $D/n_\mathrm{b}^{1/3}$, we immediately find
\begin{equation} \label{afuyiD}
a_{-1}=D\left(\frac{3\pi^2}{N_\mathrm{f}}\right)^{1/3}.
\end{equation}

At higher density, the second term in Eq.~(\ref{mInu}) dominates.
A little later, we show that the coefficient $a_1$ runs
with the strong coupling constant $\alpha\equiv\alpha_s/\pi=g^2/4\pi^2$
according to the equation
\begin{equation} \label{a1-alpha}
\sqrt{1+a_1^2}\left(1+\frac{a_1^2}{2}\right)
 -\frac{a_1^4}{2}\ln\frac{1+\sqrt{1+a_1^2}}{a_1}
=(1-2\alpha)^{-1/3}.
\end{equation}

In fact, we can prove Eq.~(\ref{a1-alpha}) by comparing
the present model at higher density with
the perturbation results.

There are several expressions for the pressure of a cold quark plasma,
e.g., those from the hard-thermal-loop perturbation theory \cite{Baier2000}
and from the weak-coupling expansion \cite{Freedman1978,Baluni1978,Toimela1985}.
Although they are different in
higher orders, the leading term is identical, i.e.,
\begin{eqnarray}
P&=& \frac{N_\mathrm{f}\mu^4}{4\pi^2}(1-2\alpha),  \label{Pqcd} \\
n_\mathrm{b} &=& \frac{N_\mathrm{f}\mu^3}{3\pi^2}(1-2\alpha). \label{nqcd}
\end{eqnarray}

At high density, because of the weak chemical equilibrium condition
$\mu_u+\mu_e=\mu_d=\mu_s$ and the quark current masses
being unimportant, we consider the flavor-symmetric case
in the present model when $m_{u0}=m_{d0}=m_{s0}=0$,
$\mu_u=\mu_d=\mu_s\equiv\mu$, $\nu_u=\nu_d=\nu_s=\nu$, and $m_i=m_\mathrm{I}$.
At high density, the second term on the right-hand side of Eq.~(\ref{mInu})
gives $m_\mathrm{I}=a_1\nu$. Accordingly, we have
$\mu^*=\sqrt{\nu^2+m_\mathrm{I}^2}=\sqrt{1+a_1^2}\nu$, and
$
{\mathrm{d}m_{\mathrm{I}}}/{\mathrm{d}\nu}
=a_1+\nu{\mathrm{d}a_1}/{\mathrm{d}\nu}
\approx a_1\
$.
(The second term is of higher order in the coupling, and can thus be ignored.)
On application of these facts, Eq.~(\ref{mumustar}) gives
the relation between the actual and effective
chemical potentials as
\begin{equation} \label{mustarsymm}
\mu^*
=\mu
 \left(
  1+\frac{a_1^2}{2}-\frac{a_1^4}{2\sqrt{1+a_1^2}}\ln\frac{1+\sqrt{1+a_1^2}}{a_1}
 \right)^{-1},
\end{equation}
and the Fermi momentum is then
\begin{equation} \label{numusymm}
\nu=\mu
\left[
 \sqrt{1+a_1^2}\left(1+\frac{a_1^2}{2}\right)
 -\frac{a_1^4}{2}\ln\frac{1+\sqrt{1+a_1^2}}{a_1}
\right]^{-1}.
\end{equation}

Substituting Eq.~(\ref{numusymm}) into the second equality of Eq.~(\ref{nunb})
and then comparing with Eq.~(\ref{nqcd}), or, equivalently,
comparing Eq.~(\ref{PT0}) with Eq.~(\ref{Pqcd}) after using
Eqs.~(\ref{mustarsymm}) and (\ref{numusymm}) etc,
we immediately obtain Eq.~(\ref{a1-alpha}).

\begin{figure}
\centering
\includegraphics[width=8cm]{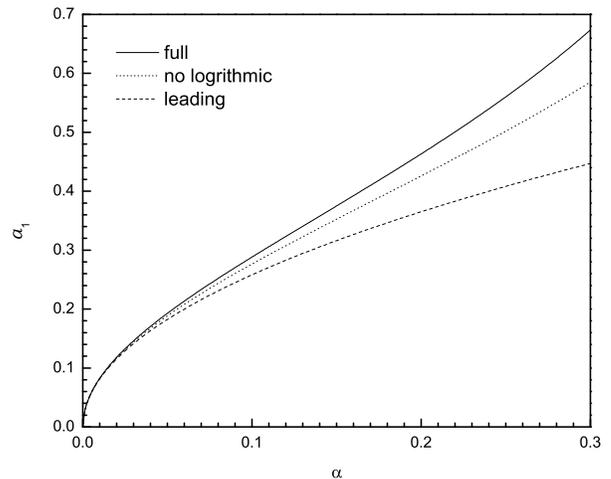}
\caption{Variation of the perturbation parameter $a_1$ with the strong coupling
$\alpha=\alpha_\mathrm{s}/\pi$. The solid curve is solved
from Eq.~(\ref{a1-alpha}), the dotted curve is calculated
by Eq.~(\ref{a1noln}) which ignores the logarithmic term
in Eq.~(\ref{a1-alpha}),
and the dashed line is the leading contribution given
by Eq.~(\ref{a1leading}).}
\label{a1alpha}
\end{figure}

Equation (\ref{a1-alpha}) determines $a_1$ as a function of
the strong coupling $\alpha$.
Although the numerical calculation of the function
is easy, an explicit expression of the functional form may be helpful.
For this we can ignore the logarithmic term that
is of the fourth order in $a_1$, giving
\begin{equation} \label{a1noln}
a_1
=\sqrt{
  \left(
    \frac{\sqrt[3]{a+\sqrt{a^2+27}}}{3}
   -\frac{1}{\sqrt[3]{a+\sqrt{a^2+27}}}
  \right)^2-1
      }
\end{equation}
with
$
a=27(1-2\alpha)^{-1/3}.\
$
Or, for simplicity, we just keep the leading term as
\begin{equation} \label{a1leading}
a_1=\sqrt{\frac{2}{3}\alpha}.
\end{equation}

The functions $a_1(\alpha)$ from Eqs.~(\ref{a1-alpha}),
(\ref{a1noln}), and (\ref{a1leading}) are given,
respectively, by solid, dotted, and dashed curves in Fig.~\ref{a1alpha}.

Now let us rewrite the quark equivalent mass as
\begin{equation} \label{mnb}
m_i
=m_{i0}+\frac{D}{n_\mathrm{b}^{1/3}}
 +C_1a_1 n_\mathrm{b}^{1/3},
\end{equation}
which is obtained by substituting Eqs.~(\ref{nunb}) and (\ref{afuyiD})
into Eq.~(\ref{mInu}).
The factor $(3\pi^2/N_\mathrm{f})^{1/3}\equiv C_{1\mathrm{max}}$
in the term proportional to the cubic-root density has been replaced with a
parameter $C_1$.
This is because we have ignored the quark current masses
in the derivation of the mass scaling.
If the finite current quark masses were included,
one would find that the factor is smaller.
Also, other approaches might give different $C_1$ values,
for example, when one considers the one-gluon-exchange
interaction \cite{ChenSW2012}, or the isospin interaction \cite{ChuPC2013}.
Therefore, we choose $C_1$ as a phenomenological model parameter
in the range of $|C_1|<\pi^{2/3}\approx 2.145$.


We have already known that the $a_1$ in Eq.~(\ref{mnb})
depend on the running coupling $\alpha$, i.e., $a_1=a_1(\alpha)$.
Now we have to discuss how the strong coupling is running.

The running coupling satisfies the renormalization-group equation:
\begin{equation} \label{RGeq}
\frac{\mathrm{d}\alpha}{\mathrm{d}\ln u^2}
=\beta(\alpha)=\sum_i\beta_i\alpha^{i+2},
\end{equation}
where the beta functions, $\beta_i$, depend generally
on which renormalization scheme is used. In the minimum subtraction
scheme \cite{Hooft1973},
they are known to the fourth order \cite{Ritbergen1997}.
The first two beta functions, $\beta_0$ and $\beta_1$,
are independent of renormalization schemes, i.e.,
$\beta_0=11/4-N_\mathrm{f}/6$\ and
$\beta_1=51/8-19N_\mathrm{f}/24$.

Simply truncating the right-hand side of
Eq.~(\ref{RGeq}) to the first order in $\alpha$,
one can easily find an inversely logarithmic solution.
It is well known, however, this simple solution
has an obviously too-large deviation from higher-order solutions.
Recently, a fast convergent expression has been obtained by
resummation over an infinite number of known terms
into a compact form \cite{Peng2006}, and the leading contribution is
\begin{equation} \label{alphaPeng}
\alpha
=\frac{\beta_0}{\beta_0^2\ln(u^2/\Lambda^2)+\beta_1\ln\ln(u^2/\Lambda^2)},
\end{equation}
where $\Lambda$\ is a QCD scale parameter, and we take $\Lambda=325$ MeV
as determined in Ref.~\cite{Peng2006}.

Another method is to use the analytic coupling
constant in the one-loop approximation~\cite{Shirkov1997}
\begin{equation} \label{alphaAnalytic}
\alpha=\frac{1}{\beta_0}
\left[
 \frac{1}{\ln{(u^2/\Lambda^2)}}+\frac{1}{1-u^2/\Lambda^2}
\right].
\end{equation}

\begin{figure}
\centering
\includegraphics[width=8cm]{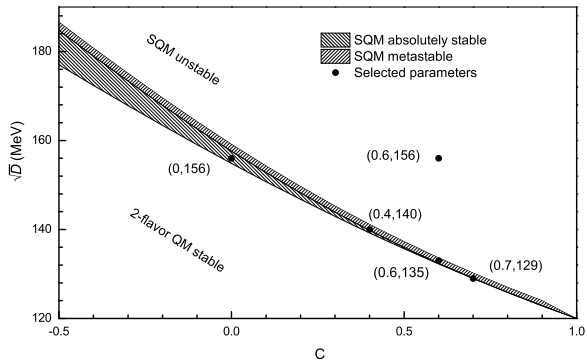}
\caption{Different range of the parameters $D$ and $C$.
Within the shaded region, SQM is stable. In the
upper-right region, SQM is unstable, while
the lower-left is the forbidden region where two-flavor quark matter
is stable.
        }
\label{figCDrange}
\end{figure}

%
%

Both Eqs.~(\ref{alphaPeng}) and (\ref{alphaAnalytic}) indicate
that the coupling runs logarithmically. The running rate
is thus much slower. Therefore, we can use an averaged constant coupling
to rewrite Eq.~(\ref{mnb}) as
\begin{equation} \label{mnbC}
m_i
=m_{i0}+\frac{D}{n_\mathrm{b}^{1/3}}
+Cn_\mathrm{b}^{1/3},
\end{equation}
where $C_1a_1$ has been grouped to be replaced
with an averaged constant $C$.

%

From Eq.~(\ref{alphaAnalytic}), it is easy to show that the maximum $\alpha$
value is $1/\beta_0$. Thus the maximum value of $a_1$ is $\sqrt{2/(3\beta_0)}$
according to Eq.~(\ref{a1leading}). Consequently, we have
\begin{equation}
C<\left(\frac{3\pi^2}{N_\mathrm{f}}\right)^{1/3}\sqrt{\frac{2}{3\beta_0}}
\approx 1.1676.
\end{equation}

In Fig.~\ref{figCDrange}, we show the different regions of the
parameters $C$ and $D$ when
taking $m_{u0}=5$ MeV, $m_{d0}=10$ MeV, and $m_{s0}=100$ MeV.
The lower-left region is forbidden where two-flavor quark matter
is stable, while in the up-right region SQM is unstable.
Only in the shaded region, SQM is absolute or metastable.
In the following calculations with Eq.~(\ref{mnbC}),
we will take the typical sets of $(C,\sqrt{D})$ pairs
where $C$ is dimensionless and $\sqrt{D}$ is in MeV:
(0.7,129), (0.6,135), (0.4,140), (0,156), (0.6,156).
These parameter pairs are indicated in Fig.~\ref{figCDrange} with solid dots.

%

We would like to emphasize that the equivalent mass
is in principle connected to the in-medium chiral
condensates \cite{Peng2003, Peng2002, Peng2005npa}, and is thus different from various effective masses.
In a NJL-type (or Schwinger-Dyson, relativistic-mean-field, $\cdots$)
description of interacting quarks, one has contributions
that affect the mass via scalar densities and the chemical
potential through vector densities.
In the present approach, an equivalent mass includes
contributions from both the scalar and vector fields.
It was explicitly shown in Ref.~\cite{Peng2003}, in the context of
symmetric nuclear matter and to leading order
in quantum hadrondynamics, that the nucleon's effective mass
involves only the scalar field $\sigma$,
while its equivalent mass is linked to both the scalar $\sigma$ field and the Lorentz vector field $\omega$.

\section{properties of strange quark matter}
\label{sec:SQM}

As usually done, we assume SQM to be composed of
up ($u$), down ($d$), and strange ($s$) quarks with
charge neutrality maintained by the inclusion of electrons ($e$) \cite{Farhi1984}.
Due to the weak interactions such as
$d,s\leftrightarrow u+e+\bar{\nu}_e$, $s+u\leftrightarrow u+d$, etc,
the chemical potentials $\mu_i$ ($i=u, d, s, e$) satisfy the
weak equilibrium conditions (neutrinos enter and leave the system freely,
and the corresponding chemical potential has been taken to be zero):
\begin{equation} \label{weakequi}
\mu_u+\mu_e=\mu_d=\mu_s.
\end{equation}

The charge neutrality condition reads
\begin{equation} \label{Chargeneut}
\frac{2}{3}n_u-\frac{1}{3}n_d-\frac{1}{3}n_s-n_e=0,
\end{equation}
with the baryon number conservation
\begin{equation} \label{Baryonconserv}
\frac{1}{3}\left(n_u + n_d + n_s\right)
=n_\mathrm{b}.
\end{equation}

\begin{figure}
\centering
\includegraphics[width=8cm]{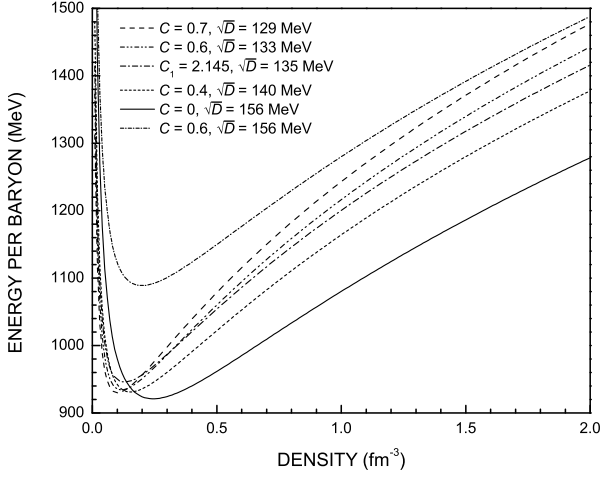}
\caption{Energy per baryon as functions of the baryon number density
for the mass scaling in Eq.~(\ref{mnbC})
at $m_{u0}=5$ MeV, $m_{d0}=10$ MeV, $m_{s0}=100$ MeV,
with different $D$ and $C$ values indicated in the legend.
\label{FigEoS}
        }
\end{figure}

In the present thermodynamic treatment,
the particle number densities $n_i\ (i=u,d,s,e)$ in
Eqs.~(\ref{Chargeneut}) and (\ref{Baryonconserv}) are connected,
by Eq.~(\ref{niT0}), to the effective chemical potentials $\mu_i^*$
which is linked to the real chemical potentials $\mu_i$
by Eq.~(\ref{mumustar}). Therefore,
Eqs.~(\ref{weakequi})-(\ref{Baryonconserv}) are four equations about
the four chemical potentials $\mu_i$ which can be solved out at given density.

According to Eq.~(\ref{mumustar}), the true and effective chemical
potentials for each flavor of quarks differ merely by a common quantity $\mu_\mathrm{I}$. Thus
the effective chemical potentials also satisfy
the similar weak equilibrium conditions:
\begin{equation}\label{mustarequi}
\mu_u^*+\mu_e=\mu_d^*=\mu_s^*.
\end{equation}
Because electrons do not participate in strong interactions,
the corresponding mass is constant. Consequently, the
effective and true chemicals potential of electrons
are the same.

We can also directly solve for the effective chemical
potentials from Eqs.~(\ref{Chargeneut})-(\ref{mustarequi}),
and then calculate all other thermodynamic quantities
from the derived expressions.

To calculate the corresponding thermodynamic quantities
such as the true chemical potentials and the pressure etc,
we need to provide the derivative of the quark mass
with respect to the density.
For Eq.~(\ref{mnbC}), it is simply
\begin{equation} \label{dnbC}
\frac{\mathrm{d}m_i}{\mathrm{d}n_\mathrm{b}}
=-\frac{D}{3n_\mathrm{b}^{4/3}}
 +\frac{C}{3n_\mathrm{b}^{2/3}}.
\end{equation}

With the quark mass scaling in Eq.~(\ref{mnbC})
and the corresponding derivative in Eq. (\ref{dnbC}),
we plot the equation of state (EOS) of SQM in Fig.~\ref{FigEoS}
for the parameters $(C,\sqrt{D})$ indicated by solid dots in Fig.~\ref{figCDrange}.
From Fig.~\ref{FigEoS} we have two observations:
(1) For parameters within the stable region, the EOS is stiffer than
those out of the region; (2) within the region,
the stiffness increases with increasing $C$ and decreasing $D$.

For the running coupling case of Eq.~(\ref{mnb}), we have
\begin{eqnarray}
\frac{\mathrm{d}m_i}{\mathrm{d}n_\mathrm{b}}
=
 -\frac{D}{3n_\mathrm{b}^{4/3}}
 +\frac{C_1}{3n_\mathrm{b}^{2/3}}
  \left[
   a_1
  +\nu\frac{\mathrm{d}a_1}{\mathrm{d}\alpha}
   \frac{\mathrm{d}\alpha}{\mathrm{d}u}
   \frac{\mathrm{d}u}{\mathrm{d}\nu}
  \right].
\label{dmidnb}
\end{eqnarray}

The derivative ${\mathrm{d}a_1}/{\mathrm{d}\alpha}$
can easily be obtained by differentiating Eq.~(\ref{a1-alpha}), giving
\begin{equation}
\frac{\mathrm{d}a_1}{\mathrm{d}\alpha}
=\frac{4(1-2\alpha)^{-4/3}}
      {
 3a_1\left(
  \frac{4+a_1^2}{\sqrt{1+a_1^2}}
  +8a_1^2\ln\frac{1+\sqrt{1+a_1^2}}{a_1}
 \right)
      }.
\end{equation}
Simply with the leading expression in Eq.~(\ref{a1leading})
we have
\begin{equation} \label{da1leading}
\frac{\mathrm{d}a_1}{\mathrm{d}\alpha}
=\frac{1}{\sqrt{6\alpha}}.
\end{equation}
Because we are trying to include first-order perturbative interaction,
we use the leading expression
in Eqs.~(\ref{a1leading}) and (\ref{da1leading})
in the numerical calculations of the present paper.

The derivative $\mbox{d}\alpha/\mbox{d}u$ in Eq.~(\ref{dmidnb})
should not be replaced with the right-hand side of Eq.~(\ref{RGeq}).
Instead, the expression is useable depending on which equation,
Eq.~(\ref{alphaPeng}) or Eq.~(\ref{alphaAnalytic}), is used.
In the former case we have
\begin{equation}
\frac{\mathrm{d}\alpha}{\mathrm{d}u}
=-\frac{\alpha^2}{u}
  \left[
   2\beta_0+\frac{\beta_1}{\beta_0\ln(u/\Lambda)}
  \right].
\end{equation}
Otherwise, if the latter is used, one then has
\begin{equation}
\frac{\mathrm{d}\alpha}{\mathrm{d}u}
=\frac{1}{\beta_0 u}
 \left[
  \frac{2}{(u/\Lambda-\Lambda/u)^2}
 -\frac{1}{2\ln^2(u/\Lambda)}
 \right].
\end{equation}

\begin{figure}
\begin{center}
\includegraphics[width=8cm]{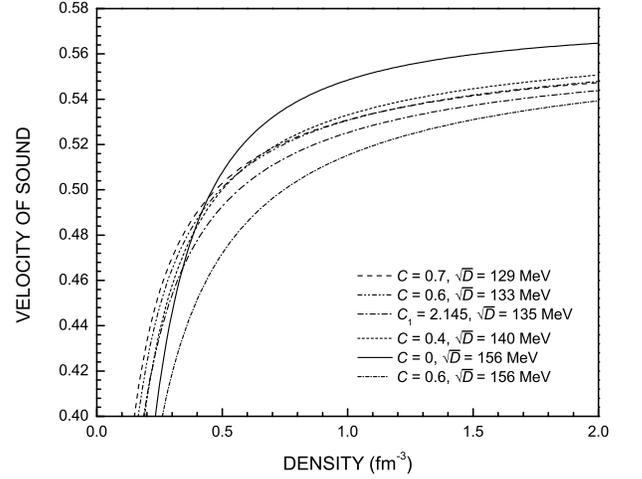}
\end{center}
\caption{The velocity of sound in SQM.}
\label{Figvos}
\end{figure}

Finally, we discuss the relation between the holistic Fermi momentum
$\nu$\ and the renormalization subtraction point $u$.
In principle, the exact relation is not available presently.
Phenomenologically we also expand it according to the Fermi momentum
$\nu$ and take to the first order as
\begin{equation}
u=c_0+c_1\nu.
\end{equation}

To use Eq.~(\ref{alphaPeng}), we have to choose
a comparatively large value for $c_0$ because it should
map the $u$ value into a reasonable perturbative range.
We take $c_0=M_\mathrm{N}=938.926$ MeV. For
analytic Eq.~(\ref{alphaAnalytic}), however,
one does not need to take care of this, and a smaller one, e.g.,
$c_0=\Lambda$, can be taken.

As for the $c_1$ value, it is generally between 2 and 3,
and we take a modest value as $c_1=2.5$.

In Fig.~\ref{FigEoS}, the dash-dot-dotted curve is for $C_1=2.145$
and $\sqrt{D}=135$ MeV with the running coupling Eq.~(\ref{alphaPeng}).

Based on the results in Fig.~\ref{FigEoS},
we can obtain the velocity of sound for SQM using 
the formula
\begin{equation} \label{vos}
v
=\sqrt{\frac{\mbox{d}P}{\mbox{d}E}}.
\end{equation}
with the results given in Fig.~\ref{Figvos}.
As the baryon number density increases, the velocity of sound also
increases. At higher densities, the curve corresponding to a larger $C$ value
approaches to the ultrarelativistic case ($1/\sqrt{3}$) more slowly.
This is understandable since there are still perturbative
interactions at higher densities.

\begin{figure}
\centering
\includegraphics[width=8cm]{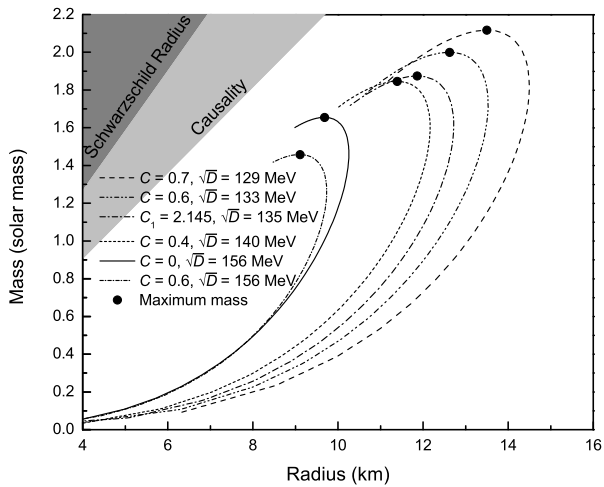}
\caption{Mass-radius relation of strange stars
for various parameter sets.
        }
\label{figMR}
\end{figure}

\section{properties of strange stars}
\label{stars}

The quark star has been an interesting subject in nuclear physics,
astrophysics, as well as in some other important fields.
Pioneer works were done with the earlier version of
the bag model  \cite{Itoh1970,Witten1984,Alcock1986,Haensel1986}.
Many further investigations have appeared in the past two decades,
such as in Refs.\ \cite{Alford2007,Weissenborn2011}.
Models other than the bag one have also been applied, e.g.
the SU(3) parity doublet model \cite{Dexheimer2013PRC},
the NJL \cite{Klahn2013}, the Komathiraj-Mahara method
\cite{Komathiraj2007,Maharaj2014}, etc.

Recently, the stellar properties were studied
with the reciprocity scaling in Eq.~(\ref{linsca}) and
the thermodynamic treatment TD-2 \cite{Torres2013}.
It is found that the maximum mass exceeds $2M_\odot$\
for all the model parameters in the whole stability window.
Therefore, this model can describe massive quark stars, but fails
to accommodate stars with low radii. As also noted by the authors,
the treatment suffers from thermodynamic inconsistency.

With the consistent thermodynamic treatment TD-3, the mass-radius relation
was previously calculated with the cubic-root scaling~\cite{Peng2000}.
It was found that the maximum mass is normally much smaller than
$2M_\odot$ \cite{Peng2000}.
Using TD-3 with the reciprocity, the case is also similar,
though it can describe stars with low radii \cite{Dexheimer2013}.
Even when one changes the confinement exponent,
which is unity in Eq.~(\ref{linsca}) and a third in Eq.~(\ref{cubicsca}),
to other values, or considering isospin interactions \cite{ChuPC2013},
the case is still similar \cite{LiA2010}.

In the preceding sections, we have obtained a new quark mass scaling
which includes both the confinement and the leading-order perturbative interactions.
On application of the scaling and the fully consistent thermodynamic treatment,
we obtain the new EOS in Fig.~\ref{FigEoS}. Let us now
apply it to solve the Tolman-Oppenheimer-Volkov equation
\begin{equation} \label{eq:TOV}
\frac{\mbox{d}P}{\mbox{d}r}
=-\frac{GmE}{r^2}
  \frac{(1+P/E)(1+4\pi r^3 P/m)} {1-2G m/r}
\end{equation}
with the subsidiary condition
\begin{equation} \label{eq:m_star}
m= \int_0^r 4\pi E r^2 \mbox{d}r.
\end{equation}
For a concise description of the solving process, one can refer to
Ref.~\cite{Peng2000}. The results are given in Fig.~\ref{figMR}
where the maximum mass is marked with full dots.
Obviously the maximum mass can be as large as $2M_\odot$.

Generally, the maximum star mass increases with increasing perturbative strength parameter $C$,
but decreases with increasing the confinement strength parameter $D$.
Therefore, going toward the lower-right direction of the stability region (shaded
in Fig.~\ref{figCDrange}) increases to as large as $2M_\odot$,
while going in the upper-right direction, it decreases to give stars with small radii,
and the maximum mass is also small, much less than $2M_\odot$.

\begin{figure*}
\centering
\includegraphics[width=13cm]{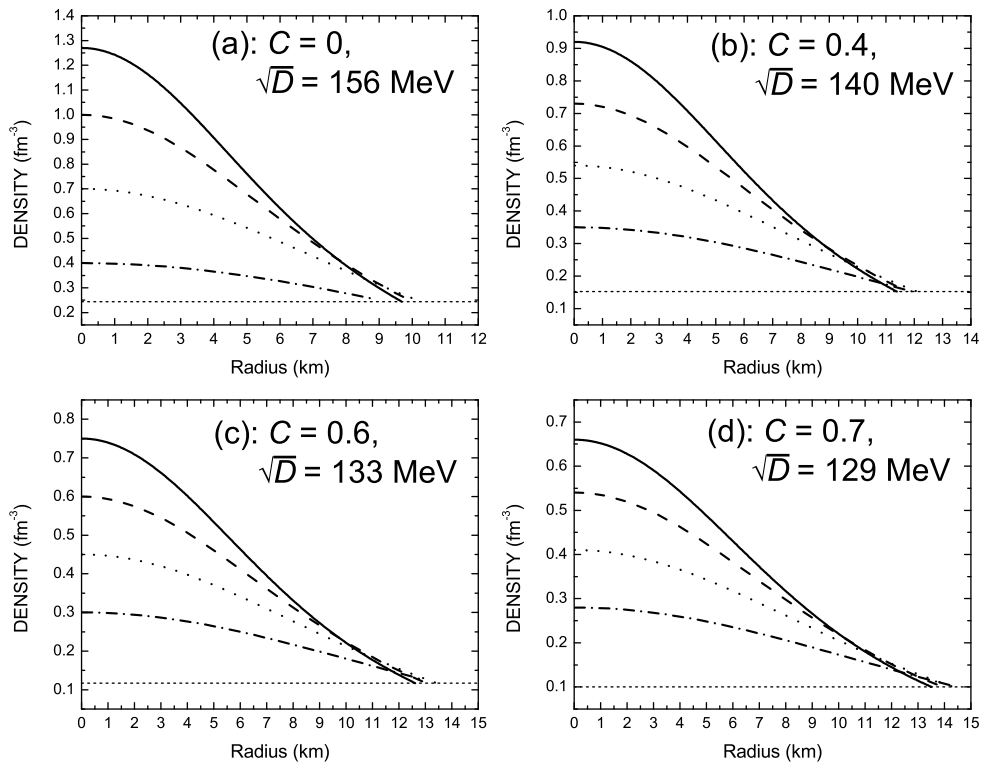}
\caption{\label{figMRnb} Density profiles for different values of the parameters $C$ and $\sqrt{D}$.
The uppermost curve on each panel is for the largest acceptable central density, while the horizontal line corresponds
to the surface density of the star.}
\end{figure*}

It should be noted that when a comparatively
large $C$ value is used to produce a large maximum mass, the
density at the surface of the star becomes very small,
even below the normal nuclear saturation density.
To show this explicitly, we draw the density profiles
as a function of the radius with four sets of the parameter
pair $(C,\sqrt{D})$. The panel (a) is for the parameter
pair ($C,\sqrt{D})=(0,156$ MeV). In this case,
the maximum star mass is smaller ($\sim$ 1.6$M_\odot$),
and the surface density is comparatively higher (0.24 fm$^{-3}$).
The panels (b), (c), and (d) are, respectively, for
(0.4, 140 MeV), (0.6, 135 MeV), (0.7, 129 MeV).
In these cases, the maximum star mass becomes larger,
but the surface density gets low, even lower than
the normal nuclear matter one. This is clearly a signal
for the phase transition to nuclear matter.
Therefore, further investigations on the QCD phase diagram \cite{Peng2008}
are necessary in future works with the new quark mass scaling.

Now we discuss a little about how to ensure the asymptotic freedom
at extremely high density. As one can see, the perturbative term in the
new quark mass scaling does not obviously decrease with increasing density.
This problem is not presently serious because, on one hand, the decreasingly
running factor $C$ ensures that the increasing velocity is slow,
and on the other hand, quarks become asymptotically free
rather slowly \cite{LeeTD2005}.
Also, if necessary, one can use a damping form factor,
to obviously ensure it, as done in Refs. \cite{ChuPC2013,ChenSW2012}
for considering the isospin or one-gluon-exchange interactions.

Furthermore, the color superconductivity \cite{Rajagopal2000}
and the strong magnetic field \cite{Chakrobarty1996,Wen2012},
 which possibly play important roles at extremely high density,
have not been considered.
The influence of these factors to the EOS of SQM, and accordingly
to the structure of compact stars, should be considered in future papers.

\section{SUMMARY}
\label{sec:sum}

We have clarified the inconsistency issues in the previous
thermodynamic treatments on SQM with density and/or temperature dependent
particle masses. We find that the fundamental differential
equation of standard thermodynamics dose not need to be modified.
Instead, the previous treatments with nonzero pressure at
the energy minimum should be discarded,
while calculations in TD-3 are still correct, and
for a full thermodynamic consistency, one just needs to
regard the original chemical potentials as effective ones.

By expanding the equivalent mass to a Laurant series and
taking the leading terms in both directions, we arrive at a new
quark mass scaling with linear confinement and leading-order
perturbative interactions. With the new quark mass scaling
and the present thermodynamics treatment, we
have studied the EOS of SQM with both constant and
running strong coupling. It is found that the new model,
which can be called an equiparticle model,
gives the EOS which can describe massive quark stars
with gravitation mass as large as 2 times the solar mass.
At the same time, it can also describe stars with
low radii, depending on the comparative strength of the confinement
and leading perturbative interactions.

\appendix

\section{Necessary conditions for the fully consistent thermodynamics of strange quark matter}
\label{appendA}

Strange quark matter is usually composed of up $u$, down $d$,
and strange $s$ quarks and electrons ($e$). Due to weak reactions,
the chemical potentials $\mu_i\ (i=u,d,s,e)$ satisfy the weak equilibrium condition,
the charge neutrality,
and the baryon number density conservation,
as in Eqs.~(\ref{weakequi})-(\ref{Baryonconserv}).
To solve out the chemical potentials $\mu_i\ (i=u,d,s,e)$ from
these equations 
for a given baryon number density $n$,
one should know the relation
between the chemical potentials $\mu_i$ and particle number densities $n_i$.
Furthermore, one should know how to calculate the system energy density
(the free energy at finite temperature), the pressure, and other
thermodynamic quantities that belong to thermodynamic treatments.
Here we derive model-independent relations any
fully consistent thermodynamic treatment for SQM must satisfy.

Let us start from the fundamental thermodynamic differential equation
by using the free energy $\bar{F}=\bar{E}-T\bar{S}$ as
\begin{equation} \label{dFbarp}
\mbox{d}\bar{F}
=-\bar{S}\mbox{d}T-P\mbox{d}V+\sum_i\mu_i\mbox{d}\bar{N}_i.
\end{equation}

For an isotropic system with homogeneously distributed particles,
it is convenient to define the corresponding intensive quantities:
the energy density $E\equiv\bar{E}/V$,
the free-energy density $F\equiv\bar{F}/V$,
the entropy density $S\equiv\bar{S}/V$,
and the particle number densities $n_i\equiv\bar{N}_i/V$.
Equation~(\ref{dFbarp}) then becomes
\begin{eqnarray}
\mbox{d}F
&=& -S\mbox{d}T +\sum_i\mu_i\mbox{d}n_i
\nonumber\\
& & +\left(-P-F+\sum_i\mu_in_i\right)\frac{\mathrm{d}V}{V}
\label{dFtem}
\end{eqnarray}
Equation (\ref{dFtem}) indicates that one should use $T$, $\{n_i\}$,
and $V$ as the independent state variables if the free-energy density $F$
is chosen as the characteristic quantity. If $F$ as a function of the independent
state variables is known, all other thermodynamic quantities can be calculated from it by
\begin{eqnarray}
S&=&
  \left.\frac{\mathrm{d}F}{\mathrm{d}T}\right|_{\{n_k\}},
 \label{appS} \\
\mu_i&=&
  \left.\frac{\mathrm{d}F}{\mathrm{d}n_i}\right|_{T,\{n_{k\neq i}\}},
  \label{appmui}\\
P&=&     -F+\sum_i\mu_in_i. 
\label{Pgen}
\end{eqnarray}

In general e.g. for strangelets, $F$ is a function of $T$, $\{n_i\}$, and $V$.
In this case, a term like $V\partial F/\partial V$ should be added to
the right-hand side of Eq.~(\ref{Pgen}). In the present case, however, we are considering
strange quark matter whose finite size effect is not significant.
The free-energy density $F$ does not depend on the volume,
and Eq.~(\ref{dFtem}) accordingly becomes
\begin{equation} \label{dF}
\mbox{d}F=-S\mbox{d}T+\sum_i\mu_i\mbox{d}n_i
\end{equation}

Because all the second-order mixed partial derivatives of an arbitrary
analytic function are equal to each other mathematically,
we can easily obtain, from Eqs.~(\ref{appS}) and (\ref{appmui}),
the relations
\begin{equation}  \label{eqDeltai}
\Delta_i\equiv
 \left.\frac{\mathrm{d}S}{\mathrm{d}n_i}\right|_{T,\{n_{k\neq i}\}}
-\left.\frac{\mathrm{d}\mu_i}{\mathrm{d}T}\right|_{T,\{n_{k}\}}=0
\end{equation}
and
\begin{equation} \label{eqDeltaij}
 \Delta_{ij}\equiv
 \left.\frac{\mathrm{d}\mu_i}{\mathrm{d}n_j}\right|_{T,\{n_{k\neq j}\}}
-\left.\frac{\mathrm{d}\mu_j}{\mathrm{d}n_i}\right|_{T,\{n_{k\neq i}\}}=0,
\end{equation}
where $i, j=u,d,s$ quarks.
Equations~(\ref{eqDeltai}) and (\ref{eqDeltaij})
are nothing but the Cauchy conditions for
the right-hand side of Eq.~(\ref{dF}) to be integrable
when the free energy is chosen as the characteristic
function.

When additional conditions are provided, e.g. these in
Eqs.~(\ref{weakequi})-(\ref{Baryonconserv}),
the free-energy density $F$ is determined as a function of
the temperature and the density $n\equiv \sum_qn_q/3$.
At a given $T$, Eq.~(\ref{dF}) gives
\begin{equation}
\frac{\mathrm{d}F}{\mathrm{d}n}
=\sum_i\mu_i\frac{\mathrm{d}n_i}{\mathrm{d}n}
\end{equation}
Therefore, we have
\begin{eqnarray}
n\frac{\mathrm{d}F}{\mathrm{d}n}
&=& n\left[
       \sum_q\mu_q\frac{\mathrm{d}n_q}{\mathrm{d}n}
      +\mu_e\frac{\mathrm{d}n_e}{\mathrm{d}n}
     \right]
\nonumber\\
&=& n\left[
       \mu\sum_q\frac{\mathrm{d}n_q}{\mathrm{d}n} -\mu_e\frac{\mathrm{d}n_u}{\mathrm{d}n}
      +\mu_e\frac{\mathrm{d}n_e}{\mathrm{d}n}
     \right]
\nonumber\\
&=&
    n\left[
       \mu\frac{\mathrm{d}}{\mathrm{d}n} \left(\sum_qn_q\right)
      -\mu_e\frac{\mathrm{d}}{\mathrm{d}n}(n_u-n_e)
     \right]
\nonumber\\
&=&
   n(3\mu-\mu_e)
  = \mu\sum_q n_q - \mu_e n
\nonumber\\
&=& \sum_q\mu_q n_q +\mu_e n_u -\mu_e n
\nonumber\\
&=& \sum_i\mu_in_i-\mu_e n_e+\mu_e n_u-\mu_e n
\nonumber\\
&=&
  \sum_i\mu_i n_i+\mu_e(n_u-n_e-n)
\nonumber\\
&=&
  \sum_i\mu_in_i,
\end{eqnarray}
where the summation on the index $i$ goes over $u, d, s, e$, while
that on $q$ goes over $u, d, s$. We have used the chemical equilibrium $\mu_u+\mu_e=\mu_d=\mu_s\equiv\mu$
and the relation
$
n_u-n_e=n
$
obtained by Eq.~(\ref{Chargeneut}) plus Eq.~(\ref{Baryonconserv}).
Considering Eq.~(\ref{Pgen}), we thus have
\begin{equation}
P=n^2\frac{\mathrm{d}}{\mathrm{d} n}\left(\frac{F}{n}\right)_T.
\end{equation}
i.e.,
\begin{equation} \label{eqDelta}
\Delta\equiv
P-n^2\frac{\mathrm{d}}{\mathrm{d} n}\left(\frac{F}{n}\right)_T=0.
\end{equation}

Equation (\ref{eqDelta}) means that the pressure at the free-energy minimum
(the energy minimum at zero temperature)
is exactly zero. In other words, the (free-)energy minimum is a mechanically
stable state. Equations~(\ref{eqDeltai}) and (\ref{eqDeltaij})
are the Cauchy conditions which ensure the existence of the system.
%
%
Therefore,
Eqs.  (\ref{eqDeltai}), (\ref{eqDeltaij}), and (\ref{eqDelta})
are the necessary conditions for any consistent thermodynamics
of strange quark matter.
One can use the pressure, (free-)energy density, chemical potentials,
and entropy given by a phenomenological model
to calculate the $\Delta$'s defined in
Eqs. (\ref{eqDeltai}), (\ref{eqDeltaij}), and (\ref{eqDelta}).
Any obvious difference from zero indicates the inconsistence degree
of the corresponding thermodynamic treatment.

\section*{ACKNOWLEDGMENTS}

The authors would like to thank the support from National Natural Science Foundation
of China (No.\ 11135011) and the key project
from Chinese Academy of Sciences (12A0A0012).

\end{document}